# Polymerisation Degree and Raman Identification of Ancient Glasses used for Jewellery, Ceramics Enamels and Mosaics.

Philippe Colomban

**Laboratoire de Dynamique, Interaction et Réactivité (LADIR), UMR 7075**

Centre National de la Recherche Scientifique & Université Pierre et Marie Curie

2 rue Henri Dunant, 94320 THIAIS, France

fax  33 1 4978 1118

colomban@glvt-cnrs.fr

**Abstract**

We demonstrate the utility of Raman spectroscopy as a technique for the identification of ancient glasses and enamel coatings of ceramics. As for any silicate glasses, the addition of network modifiers breaks the Si-O linkages and modifies the degree of polymerisation and hence the relative intensity of the Si-O bending and stretching modes. We demonstrate empirically that the ratio of these envelopes is well correlated to the glass structure and to the used firing technology. Spectral $Q^n$ components assigned to isolated and connected $SiO_4$ vibrational units allow more precise analysis. Selected porcelains, faiences, potteries and glasses representative of the different Asian, Islamic and European production technologies were studied. Modern porcelain enamels are used as compositional references.

# 1. INTRODUCTION

Many different facets of science, technology and culture are imprinted in the materials. Artefacts can be identified according to their outward appearance such as style and decoration, practical function, etc. In many cases, the identification is not easy because of their very similar appearances. On the other hand, different technologies applied to different starting materials could give very similar outward appearances from the visual point of view for products of completely different micro/nanostructure. Therefore, non-destructive and micro-probe methods are needed to examine these kinds of samples in order to understand their technology and to distinguish original artefacts from copies. The utility of Raman spectroscopy as a non-destructive technique for the characterisation of archaeological and historical artefacts has been demonstrated [1]. Bodies, glazes and pigments of many ceramics have been studied and comparison is now possible [2-6]. In rare cases, identification of synthetic or natural minerals of the ceramic body gives specific information on the technology used. Enamels (e.g. glaze coated on stoneware and porcelains, glaçures on faience and terra cotta) are glassy films, 30 to 1000µm thick, obtained by convenient firing cycles (a few days, typically). Many groups have analysed the composition (main and traces elements) of ceramic bodies in order to determine whether artefacts have similar origin. This technique is very successful for gems made of rare mineral, but fails for common minerals used as raw materials in ceramics technology. Compositional data of enamel are rare, because of the rather small thickness of enamels and their composition evolution near the surface and at the body/enamel interface. Our purpose is different. By inferring that Raman spectroscopy is able to study the structure of the glass network (previous analysis of the Si-O stretching and bending components makes the differentiation between lead-based and alkali-based compositions possible [3,4]) we will try to extract a common and quantitative fingerprint characteristic of the technique of preparation. In this paper, we will present the main parameters extracted from the Raman spectra of a large set of ancient artefacts: pottery enamels, beads, rings and mosaics tesserae from Ifriqiya [5-7], and the Islamic world (Syria, Iran, Silk Road excavations [8-10]) as well as porcelain glazes from Asia (Vietnam [2,11]) and Europe (Sèvres, St-Cloud [3,4,6,12]). After this empirical analysis step, we will apply the procedure to modern glaze samples (SINTEF Materials Technology, Norway) having well characterized compositions [13] in order to check the validity of our description. Then we will complete the analysis by comparing the $Q^n$ components of the Si-O stretching mode envelope and discuss the relationship between Raman parameters and processing techniques.

# 2. EXPERIMENTAL

Both macro- and micro-configurations Raman spectroscopy were performed on glass artefact and on the enamel surface and unglazed body of the ceramic items. The sample surface could be clearly seen under the microscope and the area of spectral examination could be well determined either on the enamel surface, at the glass/body interface or even into the body. Excitation sources were 457.9, 488, 514 and 647 nm lines from an $Ar^+$-$Kr^+$ ion laser, when using a Dilor XY spectrometer (Lille, France) and either the 532 nm from a cw frequency doubled Nd:YAG laser, or the 632.8 nm line from a HeNe laser when using a Labram Infinity Jobin-Yvon-Horiba (Longjumeau, France) instrument. The excitation powers at the samples were kept to a few milliWatts (XY) or much less (Infinity) in order to avoid any thermal effect (which should be negligible anyway in the case of the glaze as for any transparent phase). The laser beam diameters were about 100 µm for the macro-Raman (the usual 90° configuration was used) and 5 or 1.5 µm for micro-Raman measurements (backscattering configuration with x50 or x100 long focus Leitz objectives, the total magnification being 500 and 1000, respectively). Spex back illuminated CCD matrices (2000 x 256 pixels) were used for signal detection, the one on the XY Dilor spectrometer was cooled by liquid nitrogen while that used in the Infinity was cooled by Peltier effect. The spectral resolution of the XY and Raman spectrometers was $ca.$ 0.5 and 2 cm$^{-1}$, respectively. Details on the experimental set-up were described elsewhere [2-4]. To assess of the structural phase distribution over a large area of the glaze, Raman spectra were recorded at various points of each sample. This also helped to make a statistical estimation of the spectral components. Other materials science techniques such as chemical analysis, scanning electron microscopy, EDS, X-rays analysis, thermal expansion measurement and DTA were carried out to extract supplementary technical data and support Raman spectroscopic results.

In doing curve fit of Raman spectra, the baseline was first subtracted using Labspec (DILOR) software. It was assumed to be polynomial ($x^3$ or $x^4$, typically). We minimize the number of attach points in order to "isolate" at the best the two strong massifs and to attach the base line to each spectrum at about the same wavenumber position. A gaussian shape was assumed for all Raman lines because of the disordered state of examined materials. The same spectral windows were used for the extraction of the components using the Origin software peak-fitting module (Microcal Software, Inc.). We set the number of expected components (e.g. four for the stretching envelope) and assumed that all components have very close bandwidth. Nevertheless it is clear that the description is not completely univocal. The integral area under each component envelope was also calculated.

Selected ceramics were celadons and (proto)porcelains made during the 13-16$^{th}$ centuries in northern Viêt Nam [2,11], and terra cotta or faïences from Dougga, Ifriqiya (12-18$^{th}$ centuries) [5], Uzbekistan (Termez) or Pakistan (Sind) (10-16$^{th}$ centuries) [8], Iran or Syria (12-16$^{th}$ centuries) [9], Samarkand (15$^{th}$-16$^{th}$ centuries) [6] and from St-Cloud and Sèvres manufactures (18-20$^{th}$ centuries) [3, 4, 12]. Semi-cut elements of rings and diametrically pierced beads previously excavated from Carthage or Utica [7] and some mosaic elements from Roman empire [10] were considered for comparison. Finally, modern porcelain glazes [13] were used as reference.

## 3. RESULTS
### 3.1. The SiO$_4$ tetrahedron vibrational unit

Ceramic enamels and glasses are aluminosilicate networks in which the SiO$_4$ tetrahedra are joined together by the oxygen atoms located at the vertices. These SiO$_4$ tetrahedral connections are modified by the incorporation of aluminium, magnesium, iron… and alkali/alkaline earth metallic ions which changes the properties (colour, viscosity…) of glazes [14]. The depolymerization equilibrium of an initially completely linked three-dimensional tetrahedral framework of TO$_4$ tetrahedra can be express by the following equation:

O-T-O-T + M$^{v+}_{2/v}$ ⇔ 2TO$^-$ + 2/vM$^{v+}$ where M is the modifier cation of charge v+. Raman spectra obtained from the glaze could reveal these modifications of the TO$_4$ tetrahedral vibrational units (and combinations) mostly through the intensity, line-width and spectral position of the associated bands. For example, in the well-connected SiO$_4$ tetrahedral framework, as α-quartz for example, the Raman line at 464 cm$^{-1}$ originating from the bending vibration in the isolated SiO$_4$ tetrahedron ($\nu_2$ mode), is very sharp and strong (this band has the A$_{1g}$ symmetry and can also be described as the motion of the oxygen atom that bridges adjacent tetrahedra). In fused silica, this mode remains the strongest but is much broader and has various components. Briefly, one can imagine the motion of the bridging oxygen atom in a polymerised TO$_4$ network as a combination of the bending vibrations in the two adjacent TO$_4$ tetrahedron [15]. Consequently, the description in terms of symmetric stretching modes of bridging oxygen is essential to be the same as that in term of the mode originating from the bending mode of the TO$_4$ vibrational unit. In a context of molecular description where oxygen atoms belong to the stronger vibrational unit only, it is possible to analyse all the Raman spectra in the same way. By comparing Raman spectra of glassy or crystalline pure silicates and aluminates, it appears clearly that the Raman cross-section of Si-O bending and stretching modes is orders of magnitude stronger than that of Al-O bond counterparts. This arises from the more covalent character of the Si-O bond. Thus in a first approximation we could consider the SiO$_4$ entities as the vibrational unit.

### 3.2. Massifs relative intensity and "polymerisation" degree

Glass / melt polymerisation is actually a function of two key factors : bulk composition and the temperature of melt equilibration. The composition range of glass/enamel is very large (see selected composition in Table I) and our goal is to focus on the common process, which is related to the equilibrium temperature (the temperature at which the glass is quenched and the enamel is "frozen" on the artefact). Obviously employed techniques of preparation follow from the achievement of difference kiln/furnace temperatures, which in turn justify the use of different compositions. We can thus admit a correlation between the processing temperature and the melt composition, which is implicit in the preparation technique and should reflect in Raman spectra by conjugating the effect of temperature and composition. The intensity ratio of the two above-mentioned Raman massifs depends on the components and structures existing in the glaze. **Figure 1** compares representative spectra of different samples. A clear differentiation between the various glasses/glazes is possible, because the connectivity of the (SiO$_4$) polymeric units can be investigated through the relative intensities of Si-O stretching and bending modes, at ca. 1000 and 500 cm$^{-1}$ respectively. In (silica-rich) hard-paste glaze spectra, the strongest intensity is found for the band observed at ca. 480 cm$^{-1}$. This mode is strong for amorphous structures composed of connected SiO$_4$ tetrahedra as in crystalline cyclo- and tecto-silicates [16] or in any aluminosilicate networks promoted by the use of feldspar as fluxing material [17]. On the other hand, the highest intensity band of soft-paste porcelain glazes and Islamic glaçures is the stretching mode envelope (800-1200 cm$^{-1}$), characteristic of structures made with "isolated" and poorly connected tetrahedra. This behaviour is observed in glassy networks containing a large amount of elements such as alkali, alkali earth, lead or zinc ions, which break many Si-O links. In order to get a more quantitative view we have determined the ratio of the area of the bending and stretching envelopes by a fitting procedure (particular attention was paid to the samples whose composition has been determined by chemical or EDX analysis [7,10,13]). **Figure 2** shows the A$_{500}$/A$_{1000}$ ratio for all studied samples and allows for identifying different families (compositions of representative samples are given in Table 1). An initial plateau (A$_{500}$/A$_{1000}$ < 0.3, family #1) corresponds to most Islamic lead-containing glaçures of potteries from Dougga or to Islamic faiences with lustre. The alkaline-based, blue glaze of Samarcand brick also belongs to this group (see compositions in Table 1) as well as two of the blue glasses from Carthage excavation. The second family ($0.3 < A_{500}/A_{1000} < 0.8$) consists of 19$^{th}$ century lead based soft-paste porcelain enamels and of some Carthage glasses (blue, green and colourless). The large second plateau defines family #3 ($0.8 < A_{500}/A_{1000} < 1.1$), to which most Carthage glasses and 18$^{th}$ century soft-paste porcelain enamels belong. A clear slope change is observed at the crossover between family #4, corresponding to Vietnamese celadon Ca-based enamels and family #5, corresponding to Vietnamese porcelain enamels. Family #6 corresponds to hard-paste porcelain enamels. The largest ratio is observed for enamels with the lowest content of glass modifiers ions. It is clear that the $A_{500}/A_{1000}$ ratio is strongly correlated to the degree of polymerisation, i.e. to the processing temperature (from ca 1400°C for $A_{500}/A_{1000} \sim 7$, 1000°C for $A_{500}/A_{1000} \sim 1.3$ and ca 600°C or less, for $A_{500}/A_{1000} \sim 0.3$) and other parameters such as the viscosity at a given temperature..., which are characteristics of the elaboration techniques (kiln, thermal cycle,...). High $A_{500}/A_{1000}$ values correspond to silica-rich and (Ca, K,...)-poor compositions while low values correspond to modifier-rich and silica-poor compositions (see Table 1). **Figure 3** shows the plot of the ratio as a function of glass modifiers/network formers mole ratio: the mole ratio is calculated as $\Sigma$ [1/2 Na$_2$O+1/2 K$_2$O+CaO] / $\Sigma$ [SiO$_2$+1/2 Al$_2$O$_3$], and noted $\Sigma$ [M] / $\Sigma$ [Al] + [Si]. A clear relationship is observed. The slope change seems to be related to a change of the "ionicity" ratio $\Sigma$ [1/2 Al$_2$O$_3$] / $\Sigma$ [SiO$_2$], noted [Al]/[Si]. More work is needed to go further in the understanding of the relationship between $A_{500}/A_{1000}$ ratio and glass structure. In fact there is evidence from Raman spectroscopy [18], radial distribution function [19] and NMR [20] that divalent cations perturb the network more than the monovalent ones, and that ring size distribution depends on the [Al]/[Al + Si] ratio.

### 3.3. Spectral components analysis

Different spectral components ($Q^n$ for stretching components, $Q^{n'}$ for bending ones) of the above mentioned envelopes were assigned in the literature to the vibrations with zero ($Q^0$ or monomer, i.e. isolated SiO$_4$, ca. 800-850cm$^{-1}$), one ($Q^1$ or Si$_2$O$_7$ groups, ca. 950cm$^{-1}$), two ($Q^2$ or silicate chains, ca. 1050-1100cm$^{-1}$), and three ($Q^3$ or sheet-like region, ca. 1100cm$^{-1}$) and four ($Q^4$, SiO$_2$ and tectosilicates, ca. 1150-1250cm$^{-1}$) bridging oxygens (respectively four, three, two, one or zero non-bridging oxygens) per silica tetrahedral structure group [3,15,21]. By using the "quaternary" notation, the degree of polymerisation may be given by an expression like

$$2Q^3 = Q^2 + Q^4 \text{ or } 2Q^2 = Q^3 + Q^1$$

The « Raman » degree of polymerization could be extract from the precise measurement of the above quantities, but due to the rather strong fluorescence-related base line observed in the spectra of archaeological items, independent measure of $Q^3$ and $Q^4$ components is difficult, the $A_{500}/A_{1000}$ ratio measurement is less accurate but more expeditious. Nevertheless, the "quaternary" approach is valuable for samples belonging to only one or contiguous families.

**Figure 4** shows typical de-convolutions. $Q^0$, $Q^1$, $Q^2$ and $Q^3$-$Q^4$ stretching components peak at ca. 798, 948, 1040 and 1153 cm$^{-1}$. The $Q^0$ component centre of gravity seems rather low by comparison with crystalline homologue. The corresponding $Q^{n'}$ bending components are observed at ca. 258, 304, 472, 500 and 570cm$^{-1}$. The 472 cm$^{-1}$ component is associated to un-dissolved quartz traces. **Figure 5** compares the relative area of the $Q^0$, $Q^1$, $Q^2$ and $Q^3$-$Q^4$ stretching components for a series of modern hard-paste porcelain enamels with known composition and belonging to the set of samples used in Figure 2. Differentiation between $Q^3$ and $Q^4$ components was not determined because i) the base-line subtraction could shift such small intensity component, ii) the wavenumber shift induced by the connection of the last non-bridging oxygen is small and affected by neighbouring ions. The data have been plotted as a function of the mole ratio of network modifier oxides ($Na_2O$, $K_2O$, $CaO$…) over network formers ($SiO_2$, $Al_2O_3$). First, there is almost no change in the $Q^0$ component since it is related to the isolated $SiO_4$ unit in our description, and thus not very sensitive to the neighbouring species. Second, the $Q^2$ component varies very much in intensity, up to a ratio close to 0.17 and then decreases (**Fig. 5**). Simultaneously the intensity of $Q^1$ and $Q^3$-$Q^4$ components increases. It seems that addition of modifier ions first promotes $Si_3O_9$ and chains entities preserving the main connected framework until the 3D network dissociates in chain and sheet entities ($Q^3$ component increases).

## 4. DISCUSSION

The reliability of the description has been extensively studied in a previous paper [15]: the Raman spectra of the glaze of two celadons shards have been studied in many places, from the top glaze surface and on glaze fracture (13[th] and 14[th] century Ha Lan celadons). Spectra exhibiting more or less strong fluorescence have been considered. The centre of gravity of components called $Q^0$, $Q^1$ and $Q^2$ was determined at +- 5cm$^{-1}$. On the contrary the centres of gravity of the (not resolved) component assigned to $Q^3$ and $Q^4$ entities could not be measured precisely (only within +- 20cm$^{-1}$) The relative spectra intensity was measured at +- 5% for good spectra and up to +- 10% in the case of very low intensity spectra (high fluorescence). The same work has been made on modern porcelain glazes [13]. Spectra of these samples being fluorescence-free, the scattering of the data was strongly reduced: the wavenumber centres of gravity (for $Q^1$, $Q^2$ and $Q^3$ stretching and bending homologues) are measured with a precision greater than +- 4cm$^{-1}$ and the area of stretching components is determined at +- 3% or better. Thus, in the case of ancient samples the distribution of the results appears to be due to the fluorescence but also to some heterogeneity in the materials. More work is needed to explain this point.

**Figure 6** compares the centre of gravity of the components determined for the different families. The criteria used to specify the different families from the $A_{500}/A_{1000}$ plot remain valid. Data are scattered but four groups of $Q^n$ components are obvious. Families #1 and 2 exhibit strong $Q^1$ and $Q^2$ components and poor $Q^3$-$Q^4$ component. A strong 950 cm$^{-1}$ band can be associated with lead-based glass. Families #3 and 4 show strong $Q^3$-$Q^4$ components and medium $Q^1$ and $Q^2$. Family #6 shows a strong $Q^2$ component. Regarding the bending $Q^{n'}$ components, families #1 and 2 have a strong 560 cm$^{-1}$ ($Q^{2'}$) components, as in feldspar structures whereas family #6 members rather have a strong 460cm$^{-1}$ ($Q^{3-4'}$) component, like for quartz.

The last very important point is that $Q^1$ and $Q^2$ modes' centres of gravity is rather stable and specific for each kind of glaze.

## 5. CONCLUSION

The Raman fingerprint of (glassy) silicate-based compounds is a unique non-destructive tool to determine the glass structure. Because of its easy, low cost but empirical nature, the proposed Raman procedure may be successfully applied to a large variety of glass as encountered in the field of craft history and archaeology.

From the relative intensity of the bending and stretching envelopes the degree of polymerisation could be classified, allowing classifying the used technologies (e.g. the processing temperature) more easily than chemical composition. Spectral components of both envelopes give a more complete way to classify the samples belonging to only one or contiguous families. Thus, recognition of the authenticity of ancient ceramics and glasses through the identification of the processing technique of preparation become possible. For instance, it is clear that alkaline (Samarkand blue brick) and lead-based (Dougga potteries) Islamic enamels and blue Punic-Roman glasses were elaborated with a similar low temperature firing process. Strong similarities are found between Punic-Roman glasses and Islamic faience glaçures and some lead-based soft-paste porcelain glaze. On the other hand, the glass structure of Saint-Cloud Manufacture soft-paste glaze with cobalt–based underglaze blue décor is very similar to that of hard-paste porcelain. New data should be very useful to extend/improve the search of correlations between Raman parameters and composition or other process-related parameters such as melt equilibrium or temperature at a given viscosity.

**ACKNOWDLEGMENTS**

Special thanks are due to Dr O. Paulsen from Sintef Materials Technology for the reference samples he provided us with. Profs Nguyen Quang Liem, N. Ayed, T. Karmous, H. Binous, A. Louhichi, Mr. A. d'Albis, X. Faurel, G. March and Mrs F. Treppoz, A. Fäy-Hallé, V. Milande and I. Robert are kindly acknowledged for very fruitful collaboration. The author is also grateful to Dr L. Mazerolles for the use of his X-ray and EDX analysis facilities, Mrs A-M. Lagarde for drawing Figures and M. G. Sagon for his help during this work.

**FIGURE CAPTIONS**

**Fig.1**: Representative Raman spectra of enamels/glasses from Dougga pottery (Ifriqiya), soft-paste Sèvres porcelain, Carthage bead, soft-paste Saint-Cloud porcelain, Chu Đâu Vietnamese porcelain, modern hard-paste Norwich porcelain and hard-paste Sèvres porcelain.

**Fig. 2**: Master curve plot of the area ratio of Si-O bending ($A_{500}$) and stretching ($A_{1000}$) envelopes recorded on enamel/glass. (labels: gC, Carthage glass; Timour, Samarcand Bibi Khanum mausoleum; IDG: Islamic Dougga potteries; ITZ: Islamic Termez and Sind ceramics; SVR: Sèvres porcelains; IMRF: Islamic ceramics from Iran and Syria; VHL: Vietnamese Ha Lan celadon; VCL: Vietnamese Chu Đâu porcelains; StC: Saint-Cloud porcelains; NG modern hard-paste porcelains).

**Fig. 3**: Plots of the area ratio of the Si-O bending and stretching envelopes as a function of [glass modifier]/[network former] oxide mole composition ratio for modern hard-paste porcelain glaze from Sintef Technology (Norway). Error bars are smaller than the labels size.

**Fig. 4**: Example of spectral components extraction (modern hard-paste porcelain glazes).

**Fig. 5**: Plot of the Si-O stretching envelope components ("$Q^n$") as a function of the mole ratio [fluxing oxide composition]/[network former oxide composition] for reference hard-paste modern porcelain glazes. Error bar is given (+- 3%).

**Fig. 6**: Area of the different "$Q^n$" components plotted as a function of their centre of gravity wavenumber ($Q^0$ lozenge data is contaminated by $SnO_2$ cassiterite peak). See Fig. 5 and text for error bar.

Table I : Glass and glaze oxide composition (wt %) for samples representatives of the different families.

| Familly | Sample | $SiO_2$ | $Al_2O_3$ | $TiO_2$ | $Fe_2O_3$ | $Na_2O$ | $K_2O$ | $CaO$ | $PbO$ | $ZnO$ | $SnO_2$ |
|---|---|---|---|---|---|---|---|---|---|---|---|
| #6 | NG33 | 73.9 | 14.3 | 0.03 | 0.14 | 1.16 | 4.85 | 4.95 | * | * | * |
| #6 | NG6 | 63.2 | 17.9 | 0.05 | 0.2 | 1.37 | 5.72 | 11.46 | * | * | * |
| #5 | VHL | 59 | 27.2 | 0.35 | 2.4 | 0.5 | 2.6 | 10.7 | 0.5 | * | * |
| #3 | gC20 | 62 | 7 | 0.3 | 0.3 | 14 | * | 7 | 1 | * | * |
| #2 | gC18 | 53 | 7 | * | * | 17 | 2.6 | 10.5 | 0.2 | * | * |
| #1 | Timour | 47.5 | 18.8 | * | 3.4 | 5.1 | 1.2 | 6.2 | * | 9.6 | 4.8 |

* low value or not available.
NG 33, NG6: modern hard-paste glazes.
VHL: celadon glaze from Ha Lan, Viêtnam.
gC 20 and gC 18, glasses from Carthage.
Timour: blue glaze of a brick from Bibi Khanum mausoleum (Samarkand)

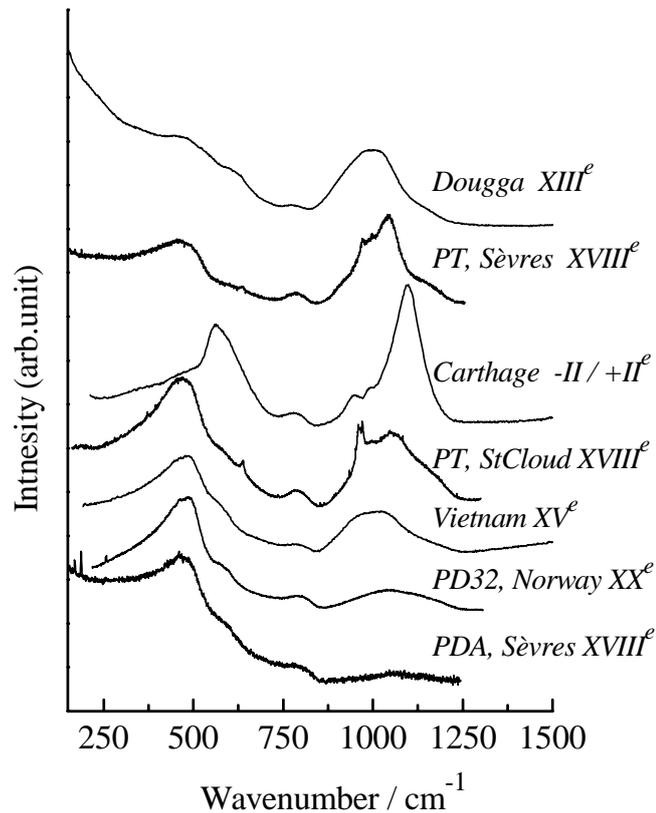

**Fig.1**: Representative Raman spectra of enamels/glasses from Dougga pottery (Ifriqiya), soft-paste Sèvres porcelain, Carthage glass bead, soft-paste Saint-Cloud porcelain, Chu Đâu Vietnamese porcelain, modern hard-paste Norwich porcelain and hard-paste Sèvres porcelain.

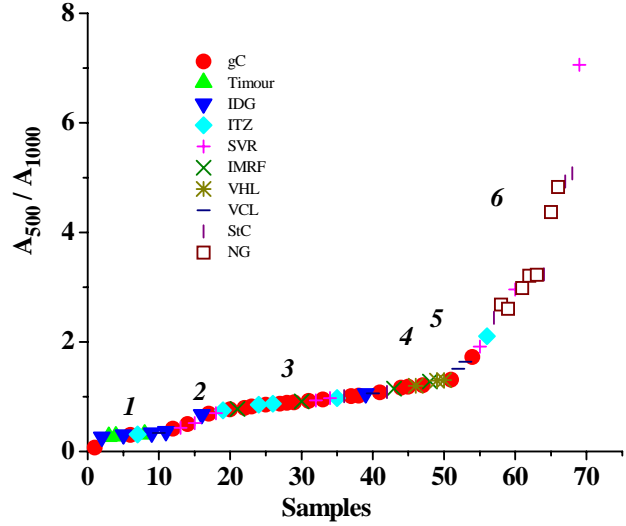

**Fig. 2**: Master curve plot of the area ratio of Si-O bending ($A_{500}$) and stretching ($A_{1000}$) envelopes recorded on enamel/glass. (labels: gC, Carthage glass; Timour, Samarkand Bibi Khanum mausoleum; IDG: Islamic Dougga potteries; ITZ: Islamic Termez and Sind ceramics; SVR: Sèvres porcelains; IMRF: Islamic ceramics from Iran and Syria; VHL: Vietnamese Ha Lan celadon; VCL: Vietnamese Chu Đâu porcelains; StC: Saint-Cloud porcelains; NG modern hard-paste porcelains).

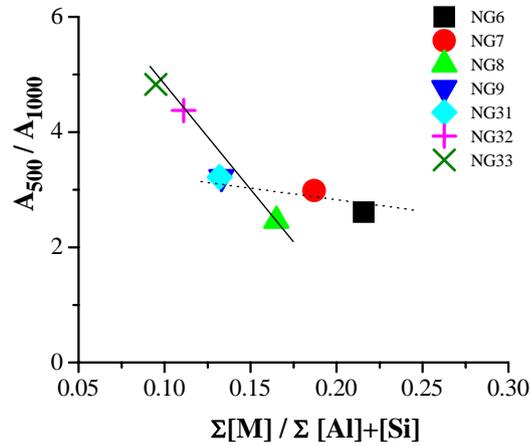

**Fig. 3**: Plots of the area ratio of the Si-O bending and stretching envelopes as a function of [fluxing oxides]/[network former oxides] mole composition ratio for modern hard-paste porcelain glaze from Sintef Technology (Norway). Error bars are smaller than the label size.

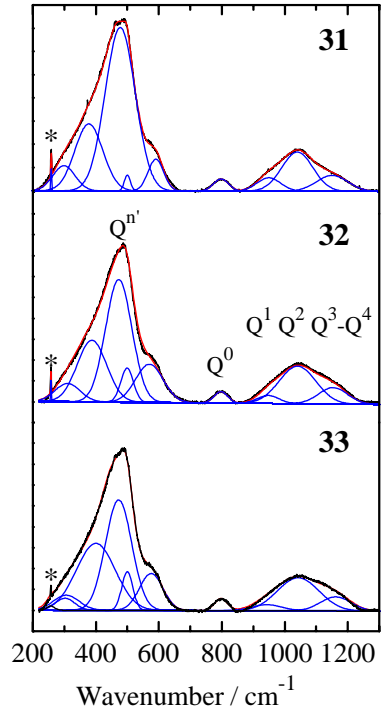

**Fig. 4**: Example of spectral components extraction (modern hard-paste porcelain glazes).

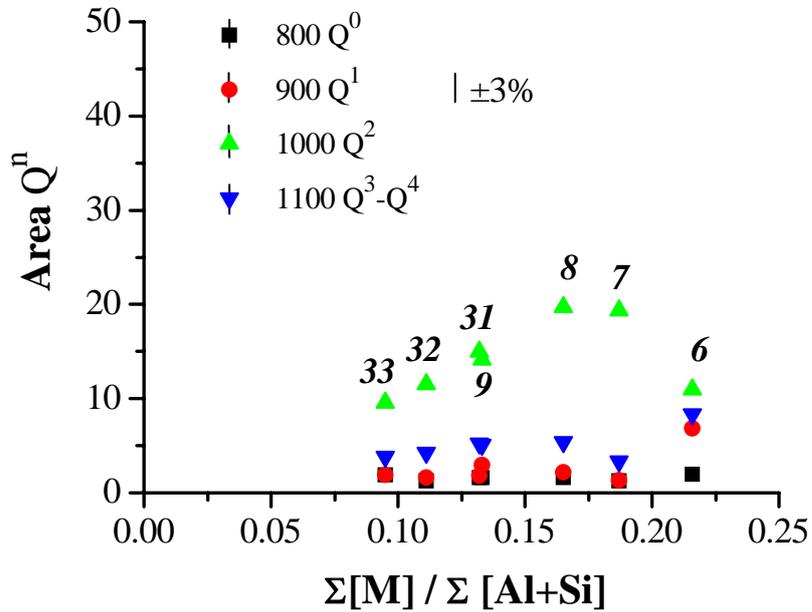

**Fig. 5**: Plot of the Si-O stretching envelope components ("$Q^n$") as a function of the mole ratio [modifier oxide composition]/[network former oxide composition] for reference hard-paste modern porcelain glazes. The error bar is given (+- 3%).

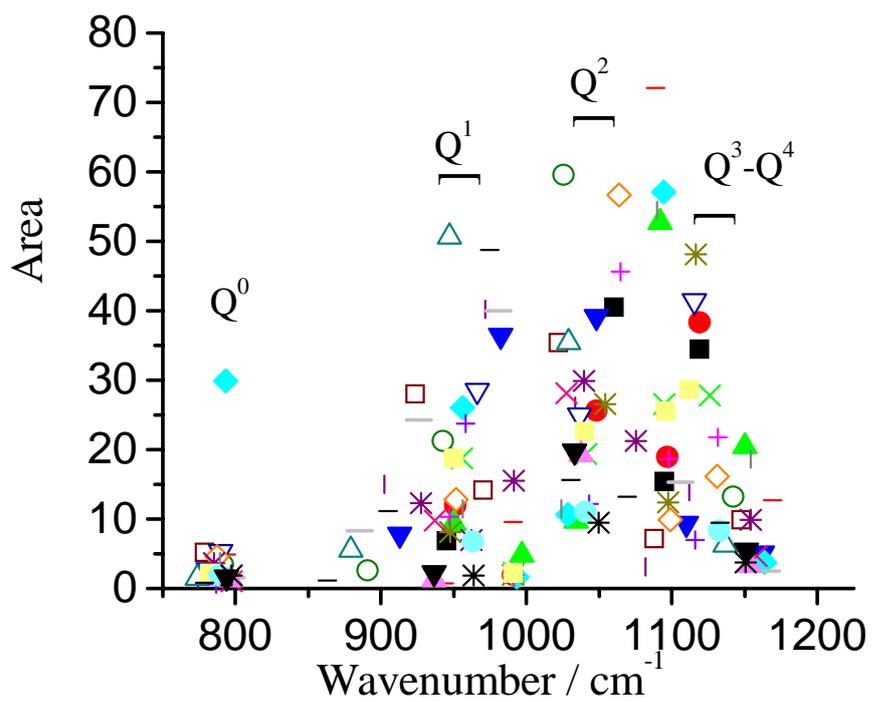

**Fig. 6:** Area of the different "$Q^n$" components plotted as a function of their centre of gravity wavenumber (see Fig 5 and text for error bar).